\documentclass[showpacs,preprint,prl]{revtex4}

\usepackage{graphicx}

\begin{document}

\title{Free Solution Electrophoresis of Homopolyelectrolytes}
\author{Pai-Yi Hsiao}
\email[E-mail: ]{pyhsiao@ess.nthu.edu.tw}
\author{Kun-Mao Wu}
\affiliation{%
    Department of Engineering and System Science, 
    National Tsing Hua University, 
    Hsinchu, Taiwan 30013, R.O.C.
}

\date{\today}

\begin{abstract}
We investigate the behavior of single polyelectrolytes in multivalent salt solutions
under the action of electric fields through computer simulations.
The chain is unfolded in a strong electric field and aligned parallel to the field 
direction, and the chain size shows a sigmoidal transition.
The unfolding electric field $E^*$ depends on the salt concentration and 
scales as $V^{-1/2}$ with $V$ being the ellipsoidal volume occupied by the chain.
The magnitude of the electrophoretic mobility of chain drastically 
increases during the unfolding. 
The fact that $E^*$ depends on the chain length provides a plausible mechanism 
to separate long charged homopolymers by size in free solution electrophoresis 
via the unfolding transition of globule polyelectrolytes condensed by multivalent salt.
\end{abstract}


\maketitle

Electrophoresis is an important separation technique in molecular biology and chemical 
analysis~\cite{viovy00,kleparnik07}. 
Nonetheless, in a free solution, polyelectrolytes (PEs) like DNA molecules cannot be size-separated 
in an electric field~\cite{olivera64} owing to the proportionality of the hydrodynamic 
friction and the total charge of the molecule to its chain length~\cite{long96,viovy00}.
Therefore, to separate homo-PEs by size is commonly carried out in a sieving matrix, 
like gel~\cite{kleparnik07}, in which biased reptation is responsible of the 
separation mechanism~\cite{cottet98,viovy00}. 
However, efforts have been continually put into free solution electrophoresis (FSE)
due to its high throughput and analysis efficiency, compared to the gel 
electrophoresis (GE). A breakthrough was done by a method called 
``end-labeled free-solution electrophoresis'' (ELFSE)~\cite{mayer94}. 
In ELFSE, the friction to charge balance is broken by attaching uncharged giant 
molecules or polymers to the ends of the PEs, producing a length-dependent electrophoretic 
mobility~\cite{ELFSE-exp}. 
Recently, Netz proposed a new strategy to separate homo-PEs by size based upon drastic
change of migration speed in the unfolding of collapsed PEs by an electric 
field~\cite{netz03}. 
He predicted that the electric field to unfold a chain, $E^*$, 
follows the scaling law $N^{-3\nu/2}$ where $N$ is the chain length and 
$\nu$ the swelling exponent.  
Therefore, longer chains will be unfolded and separated out 
using a smaller electric field.
It could be a plausible way to sequence DNA, providing a read-length of  
hundred-kilo bases~\cite{netz03}, in comparison with thousand bases 
in capillary GE and hundred bases in ELFSE~\cite{ELFSE-exp}. 
In this study, we use a coarse-grained model to verify the prediction of 
Netz for the first time by means of computer simulations.
The objective is to understand how electric fields influence the 
conformation of single chains and the electrophoretic mobility. 

There are many ways to collapse PE chains~\cite{tripathy02}, 
for e.g., reducing the temperature, 
decreasing the dielectric constant of solvent, 
modifying the chain charge density, and so on.
Adding multivalent salt into the solution is a simple way to 
initiate the collapse at ambient condition. 
It is the condensation of multivalent counterions on the chains 
which induces the chain collapse~\cite{bloomfield96}. 
The collapsed PEs exhibit the most compact structure 
at the salt concentration $C_s^*$ where the multivalent counterions  
and the chains are in charge equivalence~\cite{bloomfield96,hsiao06c}, 
resulting in a nearly zero effective chain charge and hence 
zero electrophoretic mobility~\cite{grosberg02,hsiao08a}. 
The condensation of multivalent counterions also leads to two 
very relevant phenomena, \textit{overcharging} and 
\textit{charge inversion}~\cite{grosberg02}.   
We showed in a recent study~\cite{hsiao08a} that although charge overcompensation 
(i.e.~overcharging) occurs on the surface of PE at high salt concentrations, 
the effective chain charge may not invert its sign; in other words, 
charge inversion does not necessarily happen with overcharging. 
In that study, the effective chain charge was calculated by the ratio of 
the electrophoretic mobility to the diffusive mobility, and 
the applied electric field was very weak, keeping unmodified the chain conformation.
The fact that a strong electric field could largely change the conformation and
mobility lightens up a possible way to separate PEs by size 
and drives our motivation to study the electrophoretic behavior of PEs in 
multivalent salt solutions. 
This kind of study, interplaying at the same time the strength of 
electric field and the salt concentration, is still few in literatures, 
specially by a method able to provide information at molecular level. 
It will enrich our understanding of the behavior of PEs in the presence
of multivalent counterions in a whole range of electric field. 
In Netz's study~\cite{netz03}, a PE was collapsed in a salt-free solution 
by decreasing the dielectric constant.  
Although his results gave a valuable picture of PE behavior,  
a more realistic system such as salt-collapsed PEs in electric fields should be 
investigated in detail in order to put into reality of his idea. 
Moreover,  to unravel a DNA molecule is a key issue in DNA sequencing and
diagnostics. It can be done by several methods such as optical tweezers~\cite{smith96}, 
flows~\cite{perkins95}, and electric fields~\cite{ferree03}. 
This work can provide deep insight of the unraveling of charged biopolymers 
in electric fields. 

We simulate a single PE chain using a bead-spring chain model,
consisting of $N$ monomers. 
Each monomer carries a negative unit charge $-e$ and the adjacent monomers
are connected by virtual springs described by a finitely extensible nonlinear elastic
potential
$U_{sp}(b)= -0.5 k b_{max}^2 \ln (1-(b^2/b_{max}^2))$
where $b$ is the bond length, $b_{max}$ the maximum extention, and $k$  the spring constant. 
We add (4:1)-salt into the system and hence there are three kinds of ions in the solution:
the tetravalent cations (counterions) and monovalent anions (coions) dissociated from 
the added salt, and the $N$ monovalent cations dissociated from the chain. 
The excluded volume of the monomers and ions is modeled by a shifted Lennard-Jones potential
$U_{\rm LJ}(r)= \varepsilon_{\rm LJ} \left[2(\sigma/r)^6-1\right]^2$
truncated at its minimum where $\sigma$ represents the particle diameter and
$\varepsilon_{\rm LJ}$ the coupling strength.
The particles also interact with each other via Coulomb interaction
$U_{el}(r)= k_B T \lambda_B Z_i Z_j /r$
where $k_{B}$ is the Boltzmann constant,  $T$ the absolute temperature,
$Z_i$  the charge valence, $r$ the distance and 
$\lambda_B=e^2/(4\pi \epsilon_0 \epsilon k_B T)$ the Bjerrum length
($\epsilon_0$ the vacuum permittivity).
The solvent is considered as a continuum medium 
of dielectric constant $\epsilon$ 
and its effect on the chain and ions
is implicitly taken into account by Langevin equation
$ m_i \ddot{\vec{r}}_i = - \zeta_i \dot{\vec{r}}_i +\vec{F}_c
+ Z_i e E {\hat x}+\vec{{\eta}}_i$
where $m_i$ is the $i$th particle mass and $\zeta_i$ the friction coefficient,
$\vec{F}_c=-\partial U/\partial \vec{r}_i$ the conservative force acting on 
the particle, and $\vec{\eta}_i$ the random force satisfying 
the fluctuation-dissipation theorem:
$\left< \vec{\eta}_i(t) \cdot \vec{\eta}_j(t') \right> = 6 k_B T
\zeta_i\delta_{ij} \delta(t-t')$.
The system is placed in a periodic  box,  
subject to an external electric field $E {\hat x}$ pointing 
toward $x$-direction. 
Ewald sum is applied to calculate Coulomb interaction.
We know that in dilute solutions without electric fields, the dynamics of chain 
is described by Zimm dynamics~\cite{doibook87}. 
However, under the action of electric fields, the coions and the counterions 
move in opposite directions, which largely cancels
out the hydrodynamic effect in the solutions. 
As well documented in the literatures~\cite{long96,viovy00,tanaka}, hydrodynamic 
interaction is shielded in a typical electrophoretic condition. 
Therefore, we neglect it in this study. 
We remark that this effect will become important when the chain length is very 
short~\cite{stellwagen03,grass08}.  

We assume that all the monomers and ions have identical mass $m$ and diameter 
$\sigma$.
In the following text, we choose $\sigma$, $m$, $k_B T$ to be the unit of 
length, mass, and energy, respectively. Hence, the time unit will be 
$\sigma \sqrt{m/(k_BT)}$, the unit of concentration $\sigma^{-3}$, 
the unit of electric field $k_BT/(e \sigma)$, and so on.
We set $\varepsilon_{LJ}= 0.8333$, $k=5.8333$,
$b_{max}=2$, $\lambda_B=3$, $\zeta_i= 1$. 
The chain length $N$ is varied from $12$ to $384$ and the volume of 
simulation box is changed accordingly to keep constant the monomer concentration 
$C_m$ at $3\times10^{-4}$. For a short chain ($N \le 48$), we
use a cubic box whereas for a long chain ($N>48$), a rectangular 
box elongated in the field direction is used to prevent chain overlapping 
under periodic boundary condition. 
The sides of the rectangular box are equal to $54.288$  
in $y$- and $z$-directions and $54.288 N/48$ in $x$-direction. 
Salt concentration $C_s$ is varied from $0.0$ to $6\times10^{-4}$.
The strength of the electric field $E$ is changed from $0.0$ to 
$2.0$. We launch Langevin dynamics 
simulations~\cite{note-lammps} with a time step equal to  $0.005$. 
A pre-run of $5\times 10^6$ time steps is performed to 
bring the system to a steady state and a production run is then executed 
for, at least, $5\times 10^7$ time steps.

\begin{figure}[htbp] 
\begin{center}
\includegraphics[width=0.6\textwidth,angle=270]{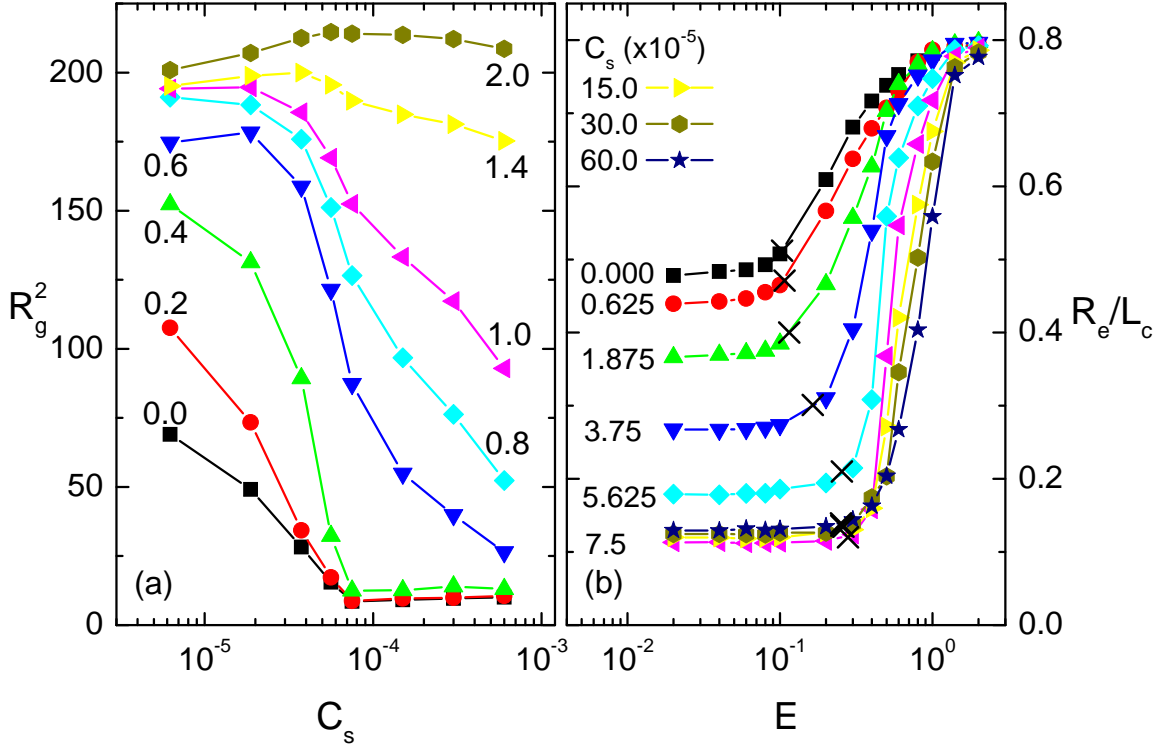}
\caption{(a) $R_g^2$ vs.~$C_s$ in different $E$.  
(b) $R_e/L_c$ vs.~$E$ at different values ($\times 10^{-5}$) 
of $C_s$.  The ``x'' symbols denote the $E^*$ at a given $C_s$ 
plotted on the associated curve.
The error of data in this Letter is either smaller than 
the symbol size of data or indicated directly by the error bar. }
\label{fig:P1N48_Rg_Re.eps}
\end{center}
\end{figure}

In Fig.~\ref{fig:P1N48_Rg_Re.eps}(a), we show the variation of chain size
for a PE of $N=48$, characterized by the square of radius of gyration, 
$R_g^2$, against $C_s$ under the action of different electric field $E$.
In zero field, the chain size decreases firstly, due to the screening or the bridging 
of tetravalent salt, up to $C_s=C_s^*=C_m/4(=7.5\times 10^{-5})$,
and then progressively increases.
This size decrease is the well-known salt-induced 
condensation~\cite{delacruz95,bloomfield96}.
Because we focus here on the behavior of condensed PEs in electric 
fields, the value of $C_s$ is not so elevated to be able to observe the 
chain decondensation.
At a really high $C_s$, the PE is expected to attain the size of its neutral 
counterpart~\cite{hsiao06a}.  
Under the action of an electric field, $R_g$ becomes larger.  
The stronger the electric field, the larger the deviation
from the chain size in the zero field. 
Moreover, there is a threshold of electric field near $E=0.4$, 
smaller than which the chain is basically not deformed by the field 
at $C_s>C_s^*$.
It shows that a coil PE (in $C_s<C_s^*$) is easier to be deformed 
than a condensed PE (in $C_s>C_s^*$).
For $E>0.4$, the chain reexpansion is suppressed by the strong field and
$R_g$ decreases with $C_s$.
A very strong field ($E=2.0$) can completely prevent the collapse of the PE 
and $R_g$ is a constant, indicating a chain unfolding.
We have verified that the unfolded chain is aligned parallel to the field direction.
The degree of unfolding can be studied by calculating 
the ratio of the end-to-end distance $R_e$ to the contour length $L_c=(N-1)b$ and
the results are plotted in Fig.~\ref{fig:P1N48_Rg_Re.eps}(b).
Each curve denotes a salt concentration and exhibits a sigmoidal variation against
$E$.  The sharp transition of the curve suggests the existence of a critical 
electric field $E^*$ to unfold a chain, over which the chain is stretched and 
$R_e/L_c$ drastically increases. 
In this study, the chain is unfolded up to a value of $R_e$, 
$80\%$ of the contour length, 
and hence displays like a rod.  
The transition occurs at different $E$, showing that $E^*$  depends on $C_s$. 
  
\begin{figure}[htbp] 
\begin{center}
\includegraphics[width=0.6\textwidth,angle=270]{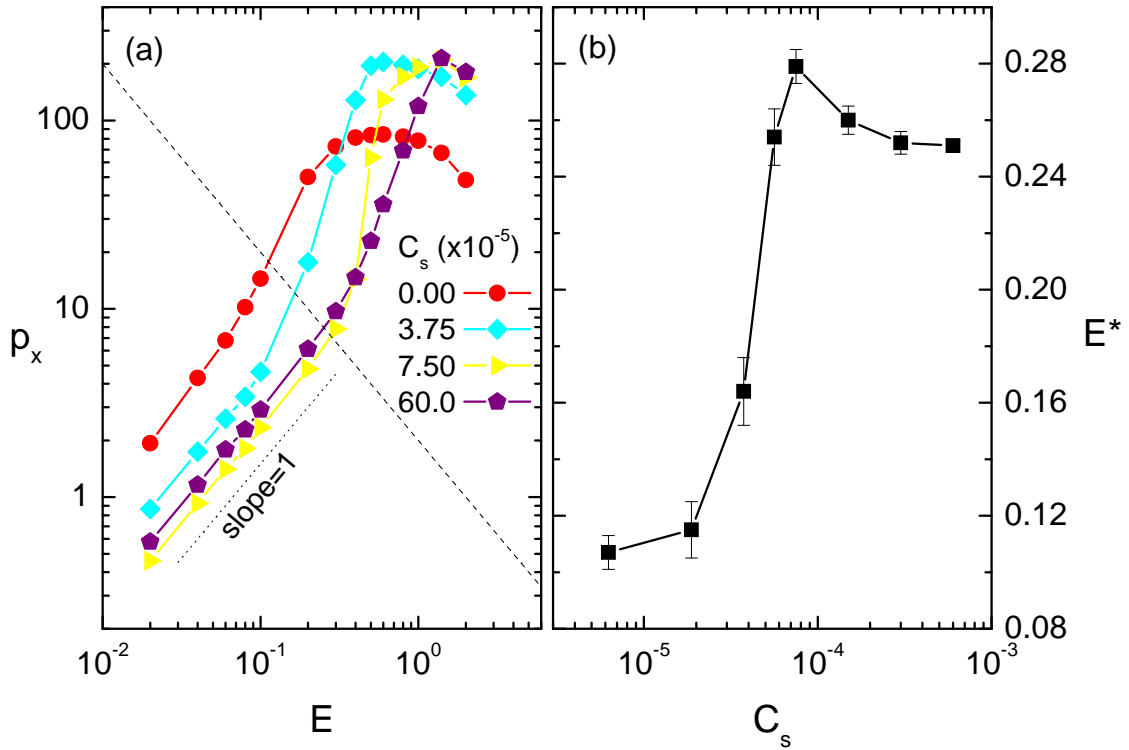}
\caption{(a) $P_x$ vs.~$E$ at different $C_s$. The dashed line denotes the 
equation $p_x=2k_BT/E$. (b) $E^*$ as a function of $C_s$.}
\label{fig:P1N48_PxE_cE_rc2.eps}
\end{center}
\end{figure}

We determine $E^*$ by equating the polarization energy 
$W_{pol}=\vec{p} \cdot \vec{E}/2$ to the thermal energy $k_BT$ 
where $\vec{p}$ is the induced dipole moment in the electric field 
$\vec{E}$~\cite{netz03}.
It is because the thermal fluctuation can destroy the polarization 
order while $W_{pol}<k_BT$ and hence, $E$ is too weak to perturb the chain 
orientation and conformation.
In solutions, charged chains can form complexes with ions. To calculate $\vec{p}$, we need to know 
first the ``members'' of the PE complex. We define the members to be the monomers 
and ions lying in a worm-shaped tube around PE, which is the union of spheres 
of radius $r_t$, centered at each center of the monomers. We choose $r_t=2$.  
Inside this tube, the kinetic energy ($3k_BT/2$) of a counterion is smaller 
than the Coulomb attraction to a monomer and hence, the ion is ``bound'' to
the chain. 
We calculate $\vec{p}$ by the equation 
$\vec{p}=\sum_{i} Z_i e(\vec{r_i}-\vec{r}_{cm})$ where $\vec{r_i}$ is the position 
vector running over all the members of the complex and
$\vec{r}_{cm}$ is the center of mass of the PE.
Fig.~\ref{fig:P1N48_PxE_cE_rc2.eps}(a) shows the $x$-component $p_x$ of $\vec{p}$ 
as a function of $E$.
We find that the linear response theory, $p_x=\alpha E$, holds true 
when $E$ is weak, and breaks down soon after $E$ surpasses $2k_BT/p_x$ 
(plotted in dashed line). 
$E^*$ is determined by the intersection of the curve with the dashed line
and is a function of $C_s$, shown in Fig.~\ref{fig:P1N48_PxE_cE_rc2.eps}(b).
$E^*$ increases with $C_s$, attains a maximum value,
and then progressively decreases. 
It shows that the most compact structure occurs at $C_s=C_s^*$. 
By plotting  $E^*$ on the corresponding curve in 
Fig.~\ref{fig:P1N48_Rg_Re.eps}(b) (with the ``x'' symbol),
we show that the $E^*$ denotes the onset where the PE starts to unfold. 
 
\begin{figure}[htbp] 
\begin{center}
\includegraphics[width=0.6\textwidth,angle=270]{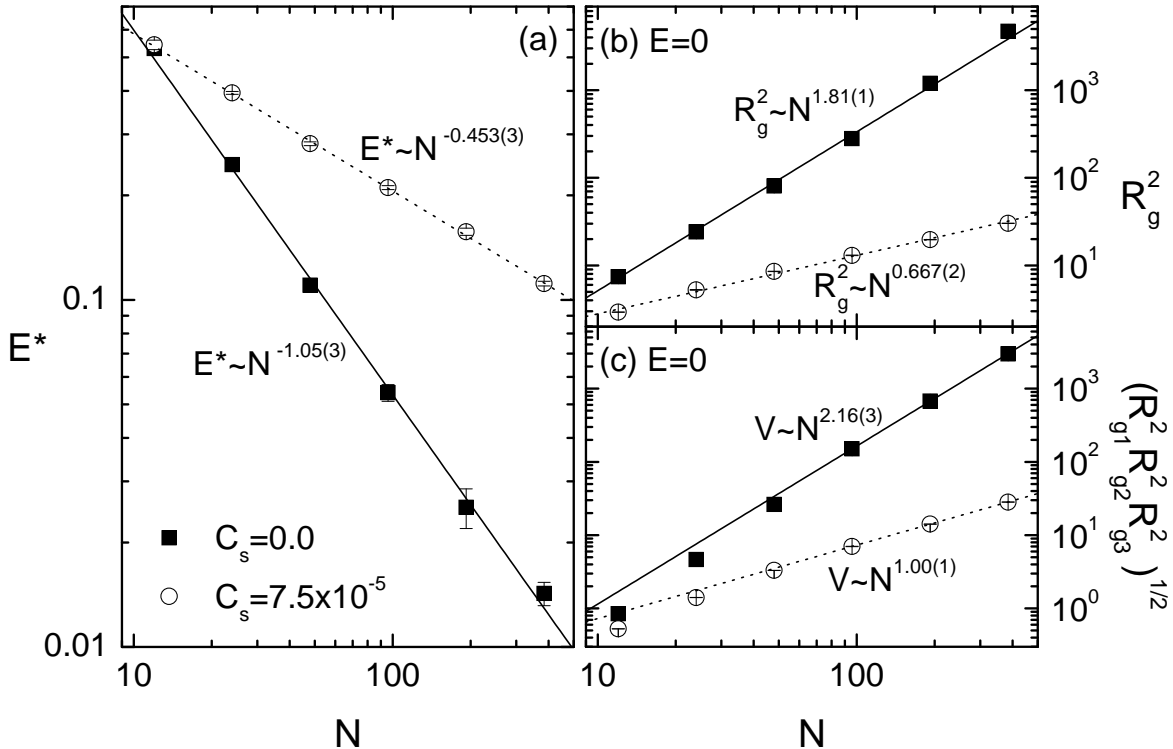}
\caption{ (a) $E^*$ vs.~$N$, (b) $R_g^2$ vs.~$N$, and 
(c) $(R_{g1}^2 R_{g2}^2 R_{g3}^2)^{1/2}$ vs.~$N$, at $C_s=0.0$ and $C_s^*$.}
\label{fig:P1Nxx_cE_Rg_V2.eps}
\end{center}
\end{figure}

Netz has predicted $E^*\sim N^{-3\nu/2}$~\cite{netz03} 
but this scaling law has
not yet been verified. Here, we study the dependence of $E^*$ on chain length 
by varying $N$ from $12$ to $384$.
We focus on two cases: $C_s=0.0$ and $C_s=C_s^*$. 
The results are presented in Fig.~\ref{fig:P1Nxx_cE_Rg_V2.eps}(a).    
In the log-log plot, $E^*$ lies on a straight line, showing a power-law
dependence. Effect of finite chain length is not important. 
Least square fits of the data (omitting the $N=12$ point) yield 
$N^{-1.05(3)}$ for $C_s=0.0$ and $N^{-0.453(3)}$ for $C_s=C_s^*$.  
According to the Netz's theory, the swelling exponent should be
$0.70(2)$ and $0.302(2)$ for the two cases, respectively.
However, a direct study of the scaling law of $R_g^2$ in the zero field 
(Fig.~\ref{fig:P1Nxx_cE_Rg_V2.eps}(b)) shows $\nu=0.91(1)$ 
for $C_s=0.0$, of which Netz's prediction has large deviation, 
and $0.334(1)$ for $C_s=C_s^*$, with which it agrees 
within $10\%$ error. We rederive here the scaling law of $E^*$. 
The general polarizability is a tensor and can be written as
$\bar{\alpha} = \epsilon_0 \bar{n}^{-1}V$ where $\bar{n}$ is the
depolarization tensor and $V$ is the volume of the dielectric
object~\cite{landau84}.  
For a fixed geometry, thus a fixed $\bar{n}$, 
the polarizability is proportional to $V$. 
In this study, $k_BT=p_xE^{*}/2=\alpha E^{*2}/2$, which implies  
$E^* \sim \alpha^{-1/2} \sim V^{-1/2}$.
In Netz's theory, $V$ was simply estimated by $R_g^3$. 
Here, we do it in a more precise way, from the gyration 
tensor of the chain. The three eigenvalues of the tensor denote 
the squares of the gyration radii, $R_{g1}^2$, $R_{g2}^2$, and
$R_{g3}^2$, for the three principal axes of inertia.
By considering the PE complex as an ellipsoid, 
we estimate $V$ as  $(R_{g1}^2 R_{g2}^2 R_{g3}^2)^{1/2}$.
Fig.~\ref{fig:P1Nxx_cE_Rg_V2.eps}(c) shows that  
$(R_{g1}^2 R_{g2}^2 R_{g3}^2)^{1/2}$ follows \textit{a priori}
a power law. The effect of finite chain length is more
important in this case because an ellipsoidal description is good
only when chain length is long.
Fitting the data for $N \ge 96$ gives $V\sim N^{2.16(3)}$ 
and $N^{1.00(1)}$ for the two cases, which yields 
$E^*\sim N^{-1.08(2)}$ and $N^{-0.50(1)}$, in good agreement
with the results of Fig.~\ref{fig:P1Nxx_cE_Rg_V2.eps}(a).  

\begin{figure}[htbp] 
\begin{center}
\includegraphics[width=0.6\textwidth,angle=270]{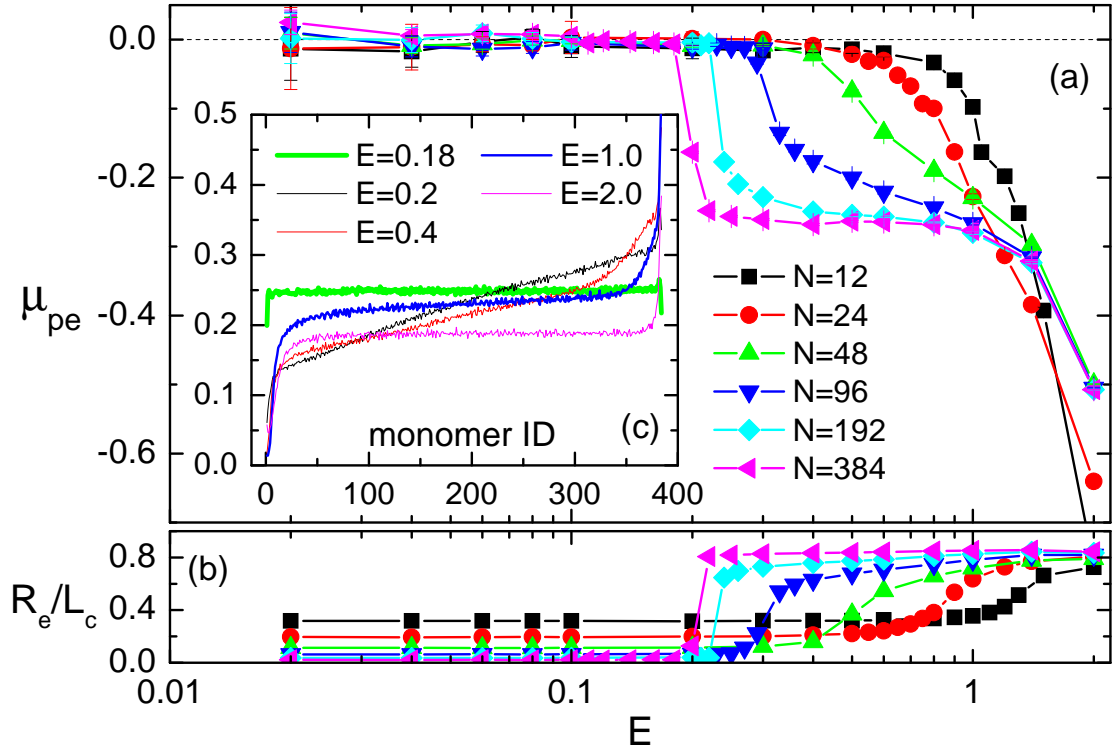}
\caption{ (a) $\mu_{pe}$ vs.~$E$ and (b) $R_e/L_c$ vs.~$E$, 
for different chain length at $C_s=C_s^*$.
(c) number of condensed tetravalent counterions on each monomer of the PE of 
$N=384$ in different strength of $E$.} 
\label{fig:P1Nxx_muPE_Re.eps}
\end{center}
\end{figure}

Knowing that $E^*$ depends on $N$, we study now the electrophoretic 
mobility of PE, $\mu_{pe}$, for different chain length. 
$\mu_{pe}$ is calculated by $v_{pe}/E$ where $v_{pe}$ is the 
velocity of the center of mass of PE in the field direction.
We focus on the case $C_s=C_s^*$ where the chain exhibits the most 
compact structure.
Fig.~\ref{fig:P1Nxx_muPE_Re.eps}(a) and (b) show, respectively, how
$\mu_{pe}$ and $R_e/L_c$ vary with $E$. 
In weak fields, $\mu_{pe}$ is zero, showing that the PEs are 
of charge neutral in the salt solutions. 
This confirms the experimental observations~\cite{bloomfield96}.
As soon as $E$ is strong enough to unfold the chain,
$\mu_{pe}$ shows a drastic decrease in accompany with chain unfolding. 
The net chain charge is hence negative. 
The longer the chain length, the sharper the unfolding transition 
and also the decrease of $\mu_{pe}$. 
The fact that the occurrence of the $\mu_{pe}$-decrease is $N$-dependent 
provides a good mechanism to separate PEs by the chain length. 
One advantage to do electrophoresis at $C_s^*$ is that only the
chains being unfolded will be moved out by the electric field, 
leaving the shorter chains in the center.
For $N\ge 96$, $\mu_{pe}$ shows a plateau  when the chain is 
completely unfolded, which is basically independent 
of chain length. This predicts a field-independent free-draining
behavior when the chains are stretched. 
We have verified that in this region, the number of the condensed 
tetravalent counterions is roughly a constant.
Fig.~\ref{fig:P1Nxx_muPE_Re.eps}(c) shows the variation of these ions
distributed along the chain backbone of $N=384$ in different $E$ fields. 
The distribution changes from a horizontal line (of value 0.25) 
at $E=0.18$ where the chain is in a collapsed state, to a line inclined toward 
the field direction at $E=0.2$ where the chain is unfolded, 
followed by a tangential ($E=0.4$) and then a horizontal ($E=1.0$) sigmoidal curve.
In a high field region $E>1$, the baseline of the distribution moves downward, 
showing a decrease of the number of the condensed ions, 
which implies the further decrease of $\mu_{pe}$. 
We have calculated the electrophoretic mobility of the condensed 
tetravalent counterions, $\mu_{+4}$. 
The value is zero when the chain is collapsed.  Once the chain is unfolded, 
$\mu_{+4}$ becomes positive and increases with $E$;
thus, the ions are not tightly bound to the chain in the electric fields
but glide along the chain following the field direction. 
They can fly away off the chain and other ions condense onto it, establishing
a dynamic equilibrium which keeps constant the number of the condensed ions.

The results clearly demonstrate the possibility to separate PEs by size 
in multivalent salt solutions under the action of appropriate electric
fields. Mapping our models to a PE in an aqueous solution at room temperature 
yields $\sigma=2.38$\AA\  and hence, the critical field $E^*=0.28$ for $N=48$ is 
equivalent to $2.9\times 10^5$V/cm in real unit.  
Therefore, to unfold a PE chain with $N=10^6$ demands 
$E^*$ of a value roughly 2 kV/cm according to the scaling law, 
which is feasible in experiments and can be applied to
separate long polymer chains in compliment to the current FSE methods.  
Our results confirm the idea of Netz~\cite{netz03}.
For practical reason, we wish the value of $E^*$ to be as small as possible.
To achieve this goal, we can use trivalent salt, or even divalent salt,
to condense PEs to a less-compact collapsed chain structure~\cite{hsiao06c}, 
which will reduce $E^*$. 
We remark that multivalent salt can also induce multi-chain aggregation.
To diminish this effect, salt of small ion size should be used as the condensing 
agent, which induces mainly single chain collapse~\cite{hsiao06a,hsiao08a}. 

In summary, we have studied the static and dynamic properties of homo-PEs in 
multivalent salt solutions under the action of electric fields.  
The electrophoretic mobility of PE shows a drastic decrease 
due to the change of the effective chain charge while the chain is 
unfolded. The unfolding electric field $E^*$ sensitively depends on
the chain length. By studying the scaling law of $E^*$, 
we presented the first, strong evidence of the feasibility to separate 
homo-PEs by size, via unfolding transition,  in free solution electrophoresis.
 
\begin{acknowledgments}
  This material is based upon work supported by the National Science Council,
  the Republic of China, under Grant No.~NSC 94-2112-M-007-023 and 
  NSC 95-2112-M-007-025-MY2.
  Computing resources are supported by the National Center for High-performance 
  Computing.
\end{acknowledgments}

\section*{Supporting Information Available}
An additional figure, which presents the electrophoretic mobility of 
condensed tetravalent counterion $\mu_{+4}$ and that of chain 
$\mu_{pe}$ as a function of $E$, is available free of
charge via the Internet at http://pubs.acs.org. 


\end{document}